\documentclass[aps,pra,superscriptaddress,reprint]{revtex4-1}
\usepackage[colorlinks=true,citecolor=blue,linkcolor=magenta]{hyperref}
\usepackage{graphicx}
\usepackage{bm}
\usepackage{dsfont}
\usepackage{amsmath,amssymb}
\usepackage{color,psfrag}
\usepackage{pstool}
\usepackage{array}
\usepackage{longtable}
\usepackage{wrapfig}
\usepackage{bbm}
\usepackage{color}

\newcommand{\ket}[1]{|#1\rangle}
\newcommand{\ex}[1]{\langle #1 \rangle}

\newcommand*{\bfrac}[2]{\genfrac{\lbrace}{\rbrace}{0pt}{}{#1}{#2}}

\usepackage{multirow}
\usepackage{lipsum}

\begin{document}
\title{Filter design for hybrid spin gates}
\author{Andreas Albrecht}
\affiliation{Institut f\"ur Theoretische Physik, Albert-Einstein-Allee 11, Universit\"at Ulm, 89069 Ulm, Germany}
\affiliation{Center for Integrated Quantum Science and Technology, Universit\"at Ulm, 89069 Ulm, Germany}
\affiliation{ICFO-Institut de Ci\`encies Fot\`oniques, 08860 Castelldefels, Barcelona, Spain.}
\author{Martin B. Plenio}
\affiliation{Institut f\"ur Theoretische Physik, Albert-Einstein-Allee 11, Universit\"at Ulm, 89069 Ulm, Germany}
\affiliation{Center for Integrated Quantum Science and Technology, Universit\"at Ulm, 89069 Ulm, Germany}

\begin{abstract}
The impact of control sequences on the environmental coupling of a quantum system can be described in terms of a filter. Here we analyze how the coherent evolution of two interacting spins subject to periodic control pulses, at the example of a nitrogen vacancy center coupled to a nuclear spin, can be described in the filter framework in both the weak and the strong coupling limit. A universal functional dependence around the filter resonances then allows for tuning the coupling type and strength. Originally limited to small rotation angles, we show how the validity range of the filter description can be extended to the long time limit by time-sliced evolution sequences. Based on that insight, the construction of tunable, noise decoupled, conditional gates composed of alternating pulse sequences is proposed. In particular such an approach can lead to a significant improvement in fidelity as compared to a strictly periodic control sequence.  Moreover we analyze the decoherence impact, the relation to the filter for classical noise known from dynamical decoupling sequences, and we outline how an alternating sequence can improve spin sensing protocols. 
\end{abstract}
\maketitle

The interaction of a quantum systems with its environment is due both to coherent control fields and environmental noise. Suppressing the environmental noise is crucial for extending the coherence time and for colored noise is frequently performed by dynamical decoupling techniques\,\cite{du09, biercuk09, delange10}. Such approaches aim at refocusing undesired couplings by means of a control field or control pulse sequence, without the need of additional ancillas or measurements. Whereas these concepts are well-understood for protecting single isolated qubits, their extension to multiple and interacting quantum systems turns out to be much more challenging\,\cite{xu12,vandersar12, gordon07}. In particular the question arises how to decouple a qubit from a noise background while at the same time retaining very specific desired interactions. Such a situation occurs in the construction of decoupled quantum gates: Even though their feasibility has been theoretically proven\,\cite{khodjasteh08, khodjasteh09, west10}, an actual implementation requires concepts that are tailored to the specifics of the system. Experimental realizations range from decoupled single\,\cite{xu12, zhang14} and multiqubit\,\cite{vandersar12, dolde13, taminiau13} spin gates to single\,\cite{timoney11} and multiqubit gates\,\cite{piltz13} in trapped ions. Another example is given by dipolar recoupling techniques in NMR experiments\,\cite{lin09, paravastu06}, which aim at the controlled and selective reintroduction of specific interactions for structural information gain.

More generally, control sequences can be used to construct and design filters\,\cite{cywinski08, kofman01, biercuk11, sousa09, gordon07} transmissive exclusively for very specific  interactions and frequencies. This insight has lead to the construction of quantum `lock-in' setups\,\cite{kotler11, cai13} or the design of Hamiltonians\,\cite{ajoy13} by choosing appropriate control parameters. Moreover decoupling sequences can be intuitively interpreted in the filter framework\,\cite{cywinski08, gordon07, uhrig07}, with decoherence arising as the overlap of a filter with the corresponding noise spectrum. Importantly, scanning a narrowband filter in frequency allows for the tomography of the noise spectrum\,\cite{alvarez11, bylander11}, the detection of ac-fields\,\cite{delange11, pham12} and even the sensing of (single) individual spins\,\cite{zhao12, taminiau12, kolkowitz12, london13, mueller14}.  Furthermore, the insight following from filter descriptions has been used for the optimization of quantum memories\,\cite{khodjasteh13} and for the construction of sensitive mass\,\cite{zhao13} and mechanical motion\,\cite{kolkowitz12mech} spectrometers. 

Here we analyze the filter description for the coherent coupling of hybrid spin systems. As an explicit example, though not limited to, we consider  the nitrogen vacancy center (NV$^-$-center) in diamond, characterized by an electronic spin-1 ground state, and its (hyperfine) coupling to individual (nuclear) spins. Both the detection\,\cite{zhao12, taminiau12, kolkowitz12, london13, mueller14} and manipulation\,\cite{taminiau12, taminiau13} of  single nuclear spins, e.g. ${}^{13}$C, ${}^{29}$Si or ${}^{14}$N nuclear spins, via the center spin have been demonstrated. 
Two relevant regimes emerge and are well-described in such a filter formalism: the weak and strong coupling limit which correspond to close\,\cite{vandersar12} and distant nuclear spins\,\cite{zhao12, taminiau12, kolkowitz12, taminiau13} from the center, respectively. 

Section\,\ref{sect1} is devoted to the filter description in the weak and the strong coupling regime, which represents a good description for short interaction times and hence small rotation angles. This description reveals universal properties around the resonance points of the filter, which allow for tuning of the interaction type and strength. Section\,\ref{sect2} then shows how to extend the applicability of these concepts beyond the short time limit and reveals the ability for tunable gate interactions based on alternating control pulse sequences. An application of this insight is provided in section\,\ref{sect3}, linking first the gate formalism to the decoherence description, followed by providing concepts for the use of the alternating sequences in spin sensing protocols with improved frequency resolution.
Appendix\,\ref{append_limit} analyzes in detail the limitations of the filter formalism. Appendix\,\ref{append_deriv} provides the derivation of the filter formulas and Appendix\,\ref{append_strong1h2} an analysis of the strong coupling limit for a spin 1/2 control spin. Last, Appendix\,\ref{append_grating} reviews the filter-grating description which simplifies the analysis and filter calculation for alternating pulse sequences.

\section{Filter formalism for coherent spin interactions}\label{sect1}
We will consider the generic Hamiltonian
\begin{equation}\label{ham1}  H=\frac{\omega}{2}\,\sigma_z+\frac{A}{2}\,\sigma_x\otimes S_z(t) \end{equation}
describing the interaction of a nuclear (`target') spin with energy separation $\omega$, hyperfine coupled with frequency $A$ to an electronic (`control') spin. Here $\sigma_k$ ($k$$\in$$\{x,y,z\}$) denotes the Pauli matrices acting on the target qubit, and for clarity the analogs on the control spin are denoted by $s_k$ in what follows. $S_z(t)$ describes a $z$-type coupling on the control spin, projected on the (pseudo)spin-1/2 submanifold as selected by a sequence of population inverting $\pi$-pulses; the latter reflected in the time dependence. Such a form holds for any type of spin couplings with significantly different energy scales which lead to the exclusion of population exchange with the control spin system. In particular we assume the control spin to be of spin-1/2 or spin-1 as in the nitrogen vacancy (NV) center in diamond and the target (nuclear) spin to be a spin-1/2 which applies to a wide variety of potential nuclear spin targets.  It is important to note that the target spin is not restricted to spin-1/2 and higher spins can be introduced straightforwardly into the formalism by replacing the spin matrices by their higher order analogs. 

For the control spin two specific configurations are considered, dependent on the two-level (sub)-system involved in the control sequence: A spin-1/2 control qubit or the spin-1 sub-manifold $\{m_s$=$\pm$1$\}$, both denoted by `\textit{spin-1/2 control spin}' in the following and describable by $S_z(t)=s_z s(t)$. A spin-1 sub-system composed of $m_s$=0 and one of the states $m_s=\pm1$, referred to as `\textit{spin-1 control}', leading to $S_z(t)=(1/2) (s_z\,s(t)\pm\mathds{1})$ with $s_z$ and $\mathds{1}$ the Pauli z-matrix and identity defined on the coupled sub-manifold, respectively. Herein the stepfunction $s(t_0)=1$ at the initial time with a sign change at each $\pi$-pulse time $t_k$, i.e. $s(t)=\sum_{k=0}^N (-1)^k \theta(t_{k+1}-t)\,\theta(t-t_k)$ for $N$-pulses with the final time $t_f\equiv t_{N+1}$ and the Heaviside stepfunction $\theta$.

The effective energy separation of an environmental spin with Larmor frequency $\boldsymbol{\omega_0}=\gamma\,\boldsymbol{B}$ ($H_0=(1/2)\boldsymbol{\omega_0}\,\boldsymbol{\sigma}$) coupled to a control spin-1 by the hyperfine interaction $\boldsymbol{A}$ ($H_{\rm hyp}=\boldsymbol{A}\boldsymbol{\sigma}\otimes 1/2(s_z s(t)\pm \mathds{1}$) is characterized by the dressed states of the unconditional (`$s_z$-independent') contributions. Thus,  $\omega=|\boldsymbol{\omega_0}\pm\boldsymbol{A}|$\,\cite{zhao14, taminiau12}, which allows for distinguishing even target spins of the same species by their position dependent coupling properties. The orthogonal hyperfine component in (\ref{ham1}) is then just the orthogonal component with respect to the dressed energy levels; parallel components can be neglected in the perturbative limit evolution considered below.  Adding a continuous resonant Rabi driving $\Omega$ on the target spin as has been performed in \cite{vandersar12}, can be described by the Hamiltonian $H=(\Omega/2)\,\sigma_x+(A/2)\,\sigma_z\otimes S_z(t)$, and thus corresponds to a global coordinate rotation of the generic form (\ref{ham1}) by replacing $\omega\sigma_z \to \Omega\sigma_x$, $A\sigma_x \to A\sigma_z$ and $\sigma_y\to-\sigma_y$. 

In the following the hyperfine coupling induced interaction is analyzed both in the strong and in the weak coupling limit under the influence of a periodic CPMG $N$-pulse control sequence\,\cite{meiboom58, vandersar12} of pulse separation $2\tau$ and total time $t=2N\tau$, i.e. $t_k=-\tau+k\,(2\tau)$ ($k$$\in$[1,N]). The effective coupling strength can then be described by a control pulse dependent filter $\mathcal{F}$ with resonances characterized by widths $1/t$ in frequency. Such a description is based on a first order Magnus expansion and holds true in the limit of small total evolution angles as shown in Appendix\,\ref{append_limit}. The extension to much longer timescales is discussed in the subsequent section. Importantly, besides allowing for specific frequency selective gate interactions, the CPMG control sequence leads to a decoherence decoupling\,\cite{delange10, naydenov11}, e.g. extends the coherence time $T_2$ for a quasi-static noise bath to\,\cite{sousa09} $T_2^{[N]}=N^{2/3}T_2$.

\subsection{Weak coupling limit $A\ll \omega$}
The weak coupling limit is essentially insensitive to the control spin being either of the `spin-1/2' or `spin-1' type, the latter requiring simply a substitution $A\to A/2$ in the subsequent expressions as a result of its effectively weaker hyperfine coupling. The total \textit{conditional} evolution in a rotating frame with respect to $H_0=(\omega/2)\sigma_z$ follows as 
\begin{equation}\label{uint_wc}  U_{\rm int}=\exp\left( -i\frac{A}{2}\,t\,\mathcal{F}_{w}(\omega\tau,N) \,\sigma_\phi\otimes s_z+\mathcal{O}([At]^2)\right) \end{equation}
with the frequency selective filter given by (see Appendix\,\ref{append_deriv})
\begin{equation}\label{filt_weak} \mathcal{F}_w(\omega\tau,N)= \frac{2}{N(\omega\tau)} \left| \frac{\sin^2(\omega\tau/2)}{\cos(\omega\tau)}\,\sin(N\omega\tau+\{\pi/2 \})  \right|  \end{equation}
and the argument in curly brackets being absent/present for $N$ even and odd, respectively.
Such a filter defines the effective coupling frequency as $(A\,\mathcal{F}_w)$ and is illustrated in figure\,\ref{b_01filter}\,(a). It reveals sharp resonances around $\omega\tau=(2k+1)\,\pi/2$ ($k\in \mathds{N}_0$) with a frequency width inversely to the total time, and a total peak height decreasing as $\propto 1/(\omega\tau)$. 
The rotation axis $\sigma_\phi$ in the resonance region
\begin{equation}\label{rescond_weak}  \omega\tau=(2k+1)\,\pi/2+c\,\pi/(2N) \end{equation}
follows a simple linear behavior, that is
\begin{equation}\label{rot_wc} \sigma_\phi=(-1)^k\,[\cos\phi \,\sigma_x-\sin\phi\, \sigma_y  ]  \quad \text{with }\phi=c\cdot (\pi/2)\,, \end{equation}
where $c\in (-2,2)$, except for $N$=1 \& $k$=0 in which case the restriction $c\in (-1,1)$ holds. This angle dependence is valid in any limit of $N$. Moreover in the limit of large $N$ ($N\gtrsim 4$), or for any $N$ at $ c$=0, the filter (\ref{filt_weak}) takes on a universal behavior around the resonances
\begin{equation}\label{filt_weak_approx} \mathcal{F}_w(c,k)\simeq\frac{4}{\pi^2}\,\frac{|\sin(c\,\pi/2)|}{|c|\,(2k+1)}  \end{equation}
resulting in the filter amplitudes $F_w(0,k)=(2/\pi) 1/(2k+1)$ for a $c$=0 conditional $\sigma_x$, and $\mathcal{F}_w(1,k)\simeq(4/\pi^2) 1/(2k+1)$ for a $c$=1 $\sigma_y$-type rotation. Importantly, the $N$ independence in (\ref{filt_weak_approx}) ensures a linear accumulation of the rotation in time. On the other hand, the $k$ dependence reflects a reduced coupling for higher order resonances. In the validity range of (\ref{filt_weak_approx}),  the total rotation angle for a given $c$ and $N$ turns out to be independent of $k$, however associated with a longer total time for increasing $k$. Moreover, the filter is symmetric in $c$ exclusively in the  limit (\ref{filt_weak_approx}) of large $N$.

\subsection{Strong coupling limit $A\gg\omega$}
\begin{figure}[tbh]
\begin{centering}
\includegraphics[scale=0.55]{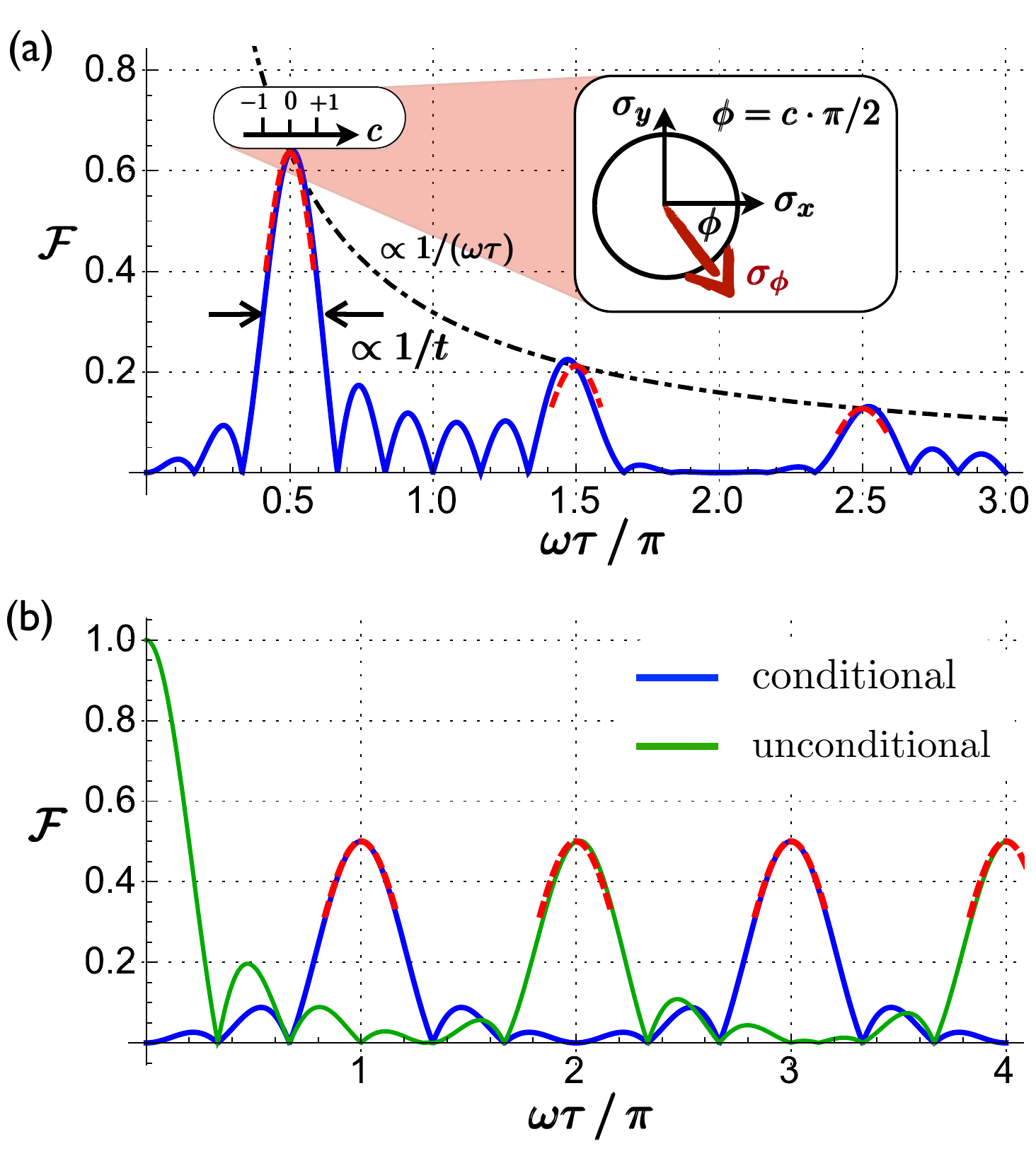}
\caption{\label{b_01filter} Gate filter for periodic control pulses. Filter function $\mathcal{F}$ describing the coherent spin interaction in the limit of small total evolution angles and for $N=6$ periodic $\pi$-CPMG-pulses with interpulse delay $2\tau$ on the control spin. \textbf{(a)} Weak coupling limit $A\ll\omega$, which leads to a purely conditional interaction as follows out of (\ref{uint_wc}).  The corresponding rotation axis $\sigma_\phi$ around the resonance peaks as given by (\ref{rot_wc}) is illustrated in the inset. Amplitudes for higher order resonances decay as $1/(\omega\,\tau)$ (black dash-dotted), whereas the resonance peak width is inversely proportional to the total time.  \textbf{(b)} Strong coupling limit $A\gg\omega$ with both conditional (blue) and unconditional (green) resonances and filters.  For both (a) and (b) red dashed lines denote the universal amplitude behavior (\ref{filt_weak_approx}) and (\ref{univ_sc}) valid in the limit of large $N$'s. }
\end{centering}
\end{figure}

In the strong coupling limit the achievable interactions crucially depend on the spin properties of the control qubit. Only for the `spin-1 configuration' a conditional interaction that is scaling linearly in time can be achieved and will be focused on in the following; the `spin-1/2 configuration' leads at most to a scalable (less relevant) unconditional  evolution and is briefly outlined in Appendix\,\ref{append_strong1h2}.\\
For a `spin-1' control qubit the evolution in a rotating frame with respect to the hyperfine Hamiltonian can be described by
\begin{equation}\label{uint_sc}\begin{split} U_{\rm int}(t)=&\exp\left(-i\,\frac{\omega}{2}t\,\left[ \mathcal{F}_{\rm c}(A\tau, N)\,\sigma_{\phi}^c\otimes s_z\right.\right.\\
&\left.\left.+\mathcal{F}_{\rm u}(A\tau, N)\,\sigma_{\phi}^u\otimes \mathds{1} \right]+\mathcal{O}([\omega t]^2)\right)  \end{split}\end{equation}
connected to the the original (\ref{ham1}) frame evolution by $U(t_n)=\exp(\mp i(A/4)t_n\,\sigma_x)\,U_{\rm int}(t_n)$ at the end of a basic decoupling sequence $t_n=t_0+2n\tau$. Upper and lower signs here and in the following are in accordance with $S_z(t)=(1/2) (s_z\,s(t)\pm\mathds{1})$, dependent on the spin-1 subsystem connected by the control pulses as previously defined. Thus the evolution is characterized by both conditional and unconditional contributions, with the corresponding filters of width $\Delta A \propto 1/t\propto 1/N$ defined as (see Appendix\,\ref{append_deriv})
\begin{eqnarray}\label{filter_sc} \mathcal{F}_{\rm c}(A\tau, N) &=&\frac{1}{2N}\Bigl| \sin\left(N \frac{A\tau}{2}+\Bigl\{\frac{\pi}{2}\Bigr\}\right) \,\tan\left(\frac{ A\tau }{2}\right)  \Bigr|  \\
\mathcal{F}_{\rm u}(A\tau, N)  &=&\frac{1}{2N}\,\left| \left[\frac{2}{A\tau}+\cot\left( \frac{A\tau}{2} \right)\right]\,\sin\left(\frac{N A\tau}{2}\right) \right| \nonumber
\end{eqnarray}
and illustrated in figure\,\ref{b_01filter}\,(b). Again the contribution in wavy brackets is present exclusively for $N$ being odd.

The resonance (region) condition, in a notation analogous to (\ref{rescond_weak}) is given by
\begin{equation}\begin{split} (A\,\tau)_c&=(2k+1)\,\pi+c\,\pi/N,\\ (A\,\tau)_u&=k\,(2\pi)+c\,\pi/N \end{split} \end{equation}
for the conditional ($c$) and unconditional ($u$) filter, respectively. The corresponding rotation axis follows as 
\begin{equation}\label{rota_sc}\begin{split} \sigma_\phi^c&=\mp\,(\cos(\phi)\sigma_z\pm\sin(\phi)\,\sigma_y) \\
			\sigma_\phi^u& = \cos(\phi)\,\sigma_z\pm\sin\phi\,\sigma_y
 \end{split}\end{equation}
with $\phi=c\,\pi/2$ for $c\in (-2,2)$, except $c\in(-1,1)$ for $\sigma_\phi^c$ and $N$=1, and restricted $k$ dependent c-ranges for $\sigma_\phi^u$ and $N=1,2$.

\begin{figure*}[!htb]
\begin{centering}
\includegraphics[scale=0.5]{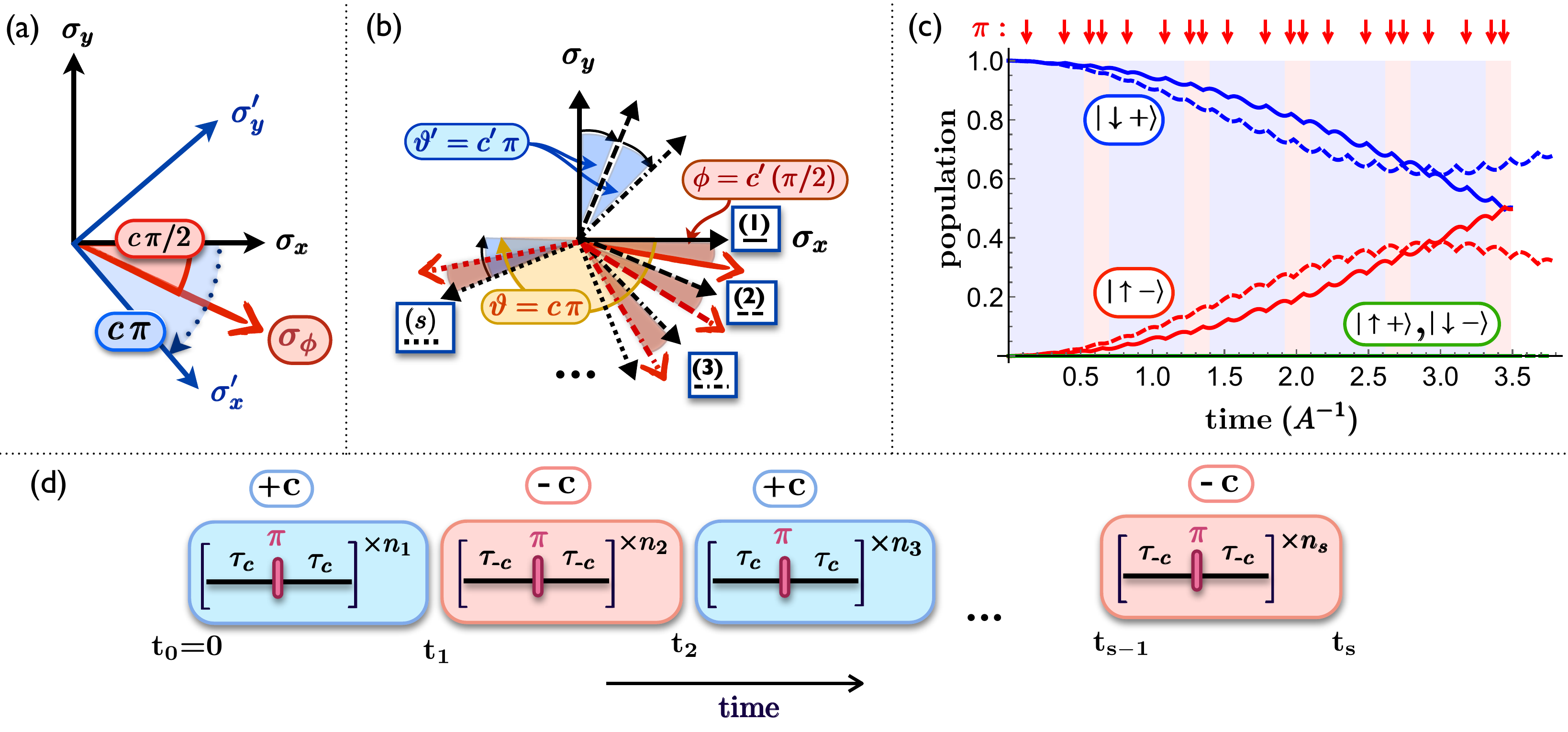}
\caption{\label{b_sliced} Sliced evolution and long time limit extension (weak coupling limit $A\ll \omega$).  \textbf{(a)} Effective coordinate system rotation by $\vartheta=c\pi$ after a single slice with even pulse number $n$, tuned to the resonance parameter $c$ (\ref{rescond_weak}) and characterized by an angle $\phi=c\,\pi/2$ of the rotation axis $\sigma_\phi$\,(\ref{rot_wc}). \textbf{(b)} Interpretation of a large $N$ sequence on resonance $c$ beyond the filter validity by decomposition into $s$ $n'$-pulse slices. Each slice is followed by a coordinate rotation $\vartheta'=c'\pi$ with $c'=c n'/N$ up to a total angle $\vartheta=c\,\pi$. Thus the rotation axis (red arrow), characterized by $\phi=c'(\pi/2)$ in the local frame, is globally rotated in time.  \textbf{(c)} Numerical simulation of a $N$=20\,$\pi$-pulse sequence for $A/\omega=0.06$, resonance order $k$=0, and the initial state $|\downarrow+\rangle$ with the control spin states $\ket{\pm}=1/\sqrt{2}\,\,(\ket{\downarrow}\pm\ket{\uparrow})$. Solid lines follow out of 10 alternating $\pm c$ ($c=1$) slices as illustrated in (d) with $n_j$$\equiv$2 $\pi$-pulses each. Dashed lines describe an equidistant $(+c)$ CPMG sequence of 20 $\pi$-pulses.  The total time shown corresponds to an expected $\pi/2$-pulse evolution based on the filter formalism. The difference in total time follows from the filter asymmetry in $c$ in the limit of small pulse numbers $n$. Red arrows indicate the $\pi$-pulse positions for the sliced (solid line) evolution; the different slices are shaded in different colors in analogy with the sequence coloring in (d). The final fidelity for the conditional $\pi/2$ evolution can be extracted as $F=99.9\%$ ($\pm c$ sliced) and $F=90\%$ (CPMG). \textbf{(d)} Alternating $\pm c$ control-pulse sequence composed of $s$ slices.        }
\end{centering}
\end{figure*}

For large $N$ ($N\gtrsim 4$) both the conditional and unconditional filter are well-approximated by the universal form 
\begin{equation}\label{univ_sc}   \mathcal{F}_{c/u}\simeq \dfrac{\sin\left( c\pi/2\right)}{c\,\pi}\,, \end{equation}
which exactly holds for any $N$ on the resonance $c$=0. Thus, $c$=0 leads to a $\sigma_z$-type rotation with $\mathcal{F}_{c/u}=1/2$, whereas $|c|$=1 corresponds to a $\sigma_y$ rotation with $\mathcal{F}_{c/u}\simeq1/\pi$. The $N$ independence of (\ref{univ_sc}) indicates the additivity in time; the additional $k$ independence reveals that the coupling amplitude, despite a modification of the ability for noise decoupling, is preserved for higher order resonances. Out of (\ref{filter_sc}) and (\ref{univ_sc}) it directly follows that the conditional filter is symmetric in $c$ in any limit of $N$, whereas this holds true for the unconditional filter exclusively for sufficiently large $N$. The (un)conditional contribution at the unconditional (conditional) resonance peak does not scale in $N$ and is thus negligible for a sufficiently large number of pulses; moreover it is zero for $c$=0 and $|c|$=1 for an even and odd number of pulses $N$, respectively, in any limit except for $N$=1\footnote{The conditional filter at the unconditional resonance is both zero for $c$=0 and $|c|$=1 for an odd number of pulses.  }.

\section{Sliced gate evolutions and alternating pulse sequences}\label{sect2}

The filter description and its associated properties discussed previously hold true as long as the interaction time and thus the evolution angle is small $\theta\ll 1$ (see Appendix\,\ref{append_limit}). For gate interactions  this forms a severe limitation, which can be overcome by alternating pulse sequences as shown in the following.

We will assume that the total evolution can be split into $s$ slices $[t_j,t_{j+1})$ ($j\in\,$[0,\,$s$-1]), with slice $j$ consisting of $n_j$-periodic CPMG-control pulses with characteristic timescale $\tau_j$ as shown in figure\,\ref{b_sliced}\,(d). Moreover $t_0$=0, the total time $t$=$t_s$ and the slice time $\Delta t_j$=$t_{j+1}$-$t_j$. Then the total evolution can be cast into the form
\begin{equation}\label{slice_evol1}U=\prod_{j=0}^{s-1} {\rm e}^{-i\,H\,\Delta t_j}={\rm e^{-i\,H_0\,t}}\,\prod_{j=0}^{s-1}{\rm e}^{i\,H_0\,t_j}U^{(j)}_{\rm int}(\Delta t_j) {\rm e}^{-i\,H_0\,t_j}   \end{equation}
where $H_0=(\omega/2)\,\sigma_z$ and $H_0=\pm (A/4)\,\sigma_x$ for the weak and strong coupling configuration, respectively.  $U_{\rm int}^{(j)}$ denotes the rotating frame evolution operator as defined in\,(\ref{uint_wc}) and (\ref{uint_sc}) with the index $(j)$ indicating that the replacement $s_z$$\to$$-s_z$ has to be made in these expressions for an odd number of preceding pulses. Assuming the rotation angle per slice to be small, its evolution is 
well-described in the filter description; e.g. for $A$$\ll$$\omega$ the rotation angle $\theta_j$=$2(A/\omega)$\,$(n_j\omega\tau_j)\,\mathcal{F}_w(\omega\tau_j, n_j)$, which is small as long as $n_j$ is. Following (\ref{slice_evol1}) the only modification consists of an effective coordinate $\exp(iH_0 \Delta t_j)$ rotation between subsequent slices $j$ and $j+1$.

The rotation of the coordinate frame is closely linked to the rotation axis angle of the evolution. This property is analyzed for the weak coupling limit in the following, however the same conclusions hold true for the strong coupling counterpart. After a single slice $j$ around a rotation axis $\phi=c\,\pi/2$ following out of conditions (\ref{rescond_weak}) and (\ref{rot_wc}), the local coordinate system can be interpreted as being rotated by 
\begin{equation}\vartheta_j=2n_j\,(\omega\tau_j)=(2k+1)n_j\pi+c\,\pi \end{equation}
for the subsequent slice. This rotation, for $n_j$ being even, is just twice the rotation axis angle $\vartheta_j=2\phi$ (see figure\,\ref{b_sliced}\,(a)). An additional $\pi$-inversion occurs for $n_j$ being odd, i.e. $\vartheta_j=2\phi+\pi$. However in that latter case also $s_z$$\to$$-s_z$ as outlined previously by the index $(j)$ to the evolution operator, and thus the rotation axis is $\pi$-inverted simultaneously and the subsequent conclusions remain unchanged\footnote{Such a coordinate frame $\pi$-inversion does not occur at the unconditional resonance in the strong coupling limit in accordance with the absence of $s_z$ in the relevant evolution operator. }. 

Thus, to globally maintain a fixed rotation axis $\phi$, the subsequent slice has to be tuned to $-c$. An alternating evolution sequence $c\to -c\to c\to -c\dots$ as depicted in figure\,\ref{b_sliced}\,(d) therefore allows for significant rotation angles beyond the perturbative limit around an arbitrary but fixed rotation axis (\ref{rot_wc}) as characterized by $\phi=c\,\pi/2$.  This is illustrated for a $\pi/2$ evolution in figure\,\ref{b_sliced}\,(c), comparing a non-alternating $N$=$20$ pulse sequence for $c$\,=\,1 to a $c$\,=\,$\pm 1$ alternating sequence of $s$=10 slices with $n_j$\,=\,$2$ pulses each.  Whereas the former simulation shows a clear deviation from the filter based expectation as a result of the small-angle approximation break-down, the latter sliced evolution leads to an (almost) perfect $\sigma_y$-type conditional $\pi/2$-rotation with fidelity $F=99.9\%$.

For an even number of alternating $\pm c$ slices (or for $c$=0), $\exp(iH_0\,t)\propto \mathds{1}$ such that the rotating frame evolution is equal to the original one. Moreover in this case and for $n_j$$\equiv$$n$, the effective `filter' for the $c$-axis rotation in analogy to\,(\ref{uint_wc}) is given by 
\begin{equation}\mathcal{F}_{\rm eff}=1/(2n)\,((c+n)\,\mathcal{F}_{\rm +c}+(-c+n)\mathcal{F}_{\rm -c})\,,\end{equation} or simply $\mathcal{F}_{\rm eff}=\mathcal{F}_{\pm c}$ whenever $\mathcal{F}$ is symmetric in $c$.

Remarkably the $c$=0 resonance peak is purely additive even in the long time limit without any adjustments, a fact that can already be seen in the $N$-independence of the corresponding resonance condition\,(\ref{rescond_weak}). On the contrary, two subsequent $c$=+1 slices would be subtractive to first order as the coordinate frame rotation inverts the global rotation axis for the second slice. 

The above insight can also be applied for the interpretation of the long time (large $N$) evolution  of equidistant CPMG sequences ($\tau_j\equiv\tau$). For that purpose, an $N$-pulse sequence characterized by a resonance parameter $c$ according to (\ref{rescond_weak}) can be formally sliced into $s$ segments of $n'=N/s$ pulses each. The resonance and rotation properties for each slice are then characterized by $c'=(n'/N)c$, as follows by comparing the resonance condition per slice $\omega \tau=(2k+1)\pi/2+c' \pi/(2n')$ to (\ref{rescond_weak}). Choosing sufficiently many slices such that the rotation per slice is small allows for the interpretation within the filter formalism. Namely, as illustrated in figure\,\ref{b_sliced}\,(b), the rotation of each slice with $c'$ is followed by a coordinate system rotation $\vartheta'=c'\pi$; up to a total rotation $\vartheta=c\pi$ for the total sequence. Essentially this equals a rotation of the rotation axis in time. Therefore, based on the slicing approach, the filter description retains its usefulness beyond the perturbative limit.

\begin{figure*}[htb]
\begin{centering}
\includegraphics[scale=0.55]{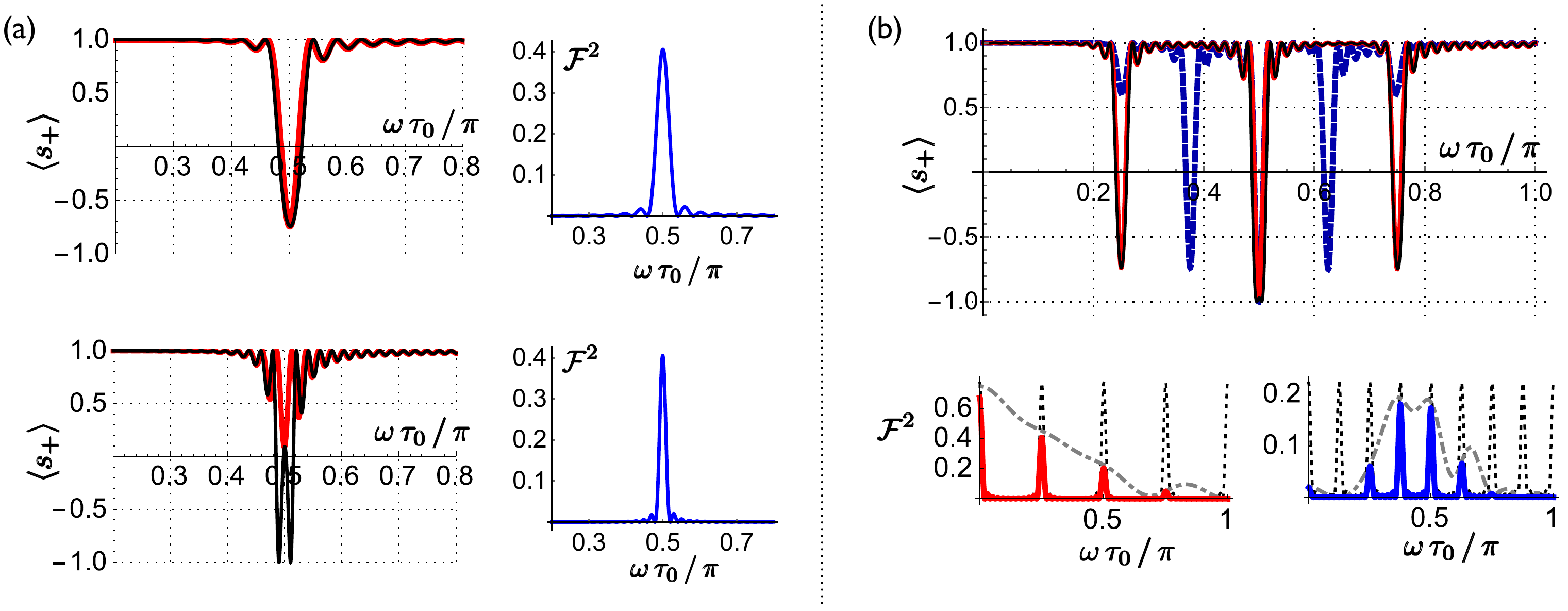}
\caption{\label{b_sensing} Spin sensing (weak coupling $A/\omega=0.05$). \textbf{(a)} Control spin coherence $\ex{s_+}$ (red) for a CPMG pulse number $N=24$ (upper) and $N=48$ (lower) with total time $t=2N\tau_0$. Black lines indicate the coherence evolution as expected out of the filter formalism (\ref{spin_sens1}). The squared filter $\mathcal{F}^2$ is illustrated for comparison.  \textbf{(b)} Coherence for an alternating $c=\pm 1$ $N=48$ pulse sequence with $M$=12, $2n$=4 (red) and $M=6$, $2n$=8 (blue, dashed). Black lines correspond to the filter expectation for the $M$=12 case. Lower panel: Filter (\ref{filtgrating}) for the $M$=12 (left) and $M$=6 (right) configurations and its composition (in arbitrary units) out of grating (dotted) and block filter (dash-dotted).     }
\end{centering}
\end{figure*}

\section{Spin sensing and frequency filter for alternating pulse sequences}\label{sect3}
The conditional coherent interaction induced by Hamiltonian (\ref{ham1}) leads to a mutual, time-periodic entanglement of the control and target spin. This effect can be observed as a coherence decay on the control spin subsystem, which along with the frequency selectivity of the control pulse sequence is widely used for sensing individual environmental spins\,\cite{zhao12, taminiau12, kolkowitz12}. More precise, the coherence follows as (with $s_+$ the ladder operator for the control spin) $\ex{s_+(t)}=\ex{s_+(0)}\,\text{tr}[ {U_{\rm int}^{(s_z=+1)\dagger}} U_{\rm int}^{(s_z=-1)} \rho_{\rm c} ]$, which in the $A\ll\omega$ limit and assuming the target initial state $\rho_{\rm c}=(1/2)\mathds{1}$ to be completely mixed, leads to
\begin{equation}\label{spin_sens1} \ex{s_+(t)}\simeq \ex{s_+(0)}\,\cos\left(At\,\mathcal{F}_w(\omega\tau,N)\right)\,.  \end{equation}

For sufficiently short times, the coherence evolution follows as $\ex{s_+(t)}\simeq \ex{s_+(0)}\,\exp(-(1/2)\,A^2\,t^2 \mathcal{F}_w(\omega\tau,N)^2)$. This latter expression is formally equivalent to the decay under classical noise\,\cite{zhao12}, with the corresponding filter $\mathcal{F}_{\rm class}=\mathcal{F}_w^{\,2}$\,\cite{cywinski08, sousa09} and assuming a noise bath with discrete spectrum $S(\omega')=\delta(\omega-\omega')$.  

The validity range of (\ref{spin_sens1}) is limited by the applicability range of the filter formalism. In practice, as simulated for a CPMG sequence in figure\,\ref{b_sensing}\,(a), the coherence is well-described by (\ref{spin_sens1}) within the initial sign inverting regime $\ex{s_+(0)}$$\to$$-\ex{s_+(0)}$, i.e. up to a rotation angle $\pi$ of the conditional evolution.  It forms the typical regime used for spin-sensing. Going beyond this timescale reduces the dip depth of the main resonance again, whereas the vicinity regions still increase in depth. This then leads to an oscillatory broadened regime associated with a loss of frequency resolution. More precise,  such a regime makes it hard to distinguish different spins of similar frequency as a result of their overlapping signal envelopes; in addition to the drawbacks associated with the (potentially) reduced coherence dip amplitude and the more intricate signal pattern. Moreover the filter formula description (\ref{spin_sens1}) breaks down at these timescales; except for $c$=0 ($\omega\tau_0$=$\pi/2$ in figure\,\ref{b_sensing}\,(a)) as is clear out of the `slicing' analysis of the preceding section. Thus there exists a trade-off between an anticipated increasing sensing resolution in time out of the filter description ($\Delta\omega \propto 1/t$) and an upper limitation by the onset of the oscillatory broadening at longer times; besides the limitation on the coherence time set by the classical noise background.

Extending the time until the onset of the oscillatory regime can be achieved by lowering the effective coupling amplitude. That way the filter description and its time inverse frequency width scaling retains its validity on a longer timescale, which, provided the (decoupled) coherence time exceeds that timescale, enables to further increase the sensing resolution\,\cite{zhao14}. Thus, for a fixed total time, appropriately tuning the coupling strength allows to turn a signal with previously oscillatory broadened envelope and reduced dip depth, into a clearly distinct peak structure of larger or maximal  peak depth, each peak width scaling with the inverse total time.    

Such an amplitude damping effect is realized by higher order resonances ($k$$>$0)\,\cite{taminiau12}, by superimposing a continuous Rabi driving on the control spin\,\cite{mkhitaryan15}, or by lowering the effective coupling amplitude by gradually reducing the periodicity of the control sequence\,\cite{zhao14}. This latter concept can be achieved by an $N$-pulse sequence composed of $s$  alternating $\pm c$\,\,slices with $n$ pulses each (figure\,\ref{b_sliced}\,(d)), choosing the inter-pulse timescales such that
\begin{equation} \tau_{\pm c}=\tau_0\pm c\,\frac{\tau_0}{(2k+1)\,n}\,.  \end{equation}
Here $\tau_0=(1/2)\,(\tau_{+c}+\tau_{-c})$ or simply $\tau_0=t/(2N)$ for an even number of slices $s$. That way, at the resonance condition $\omega\tau_0=(2k+1)\,\pi/2$, the system is alternatingly driven at the $\pm c$ resonance, i.e. $\omega\,\tau_{\pm c}=(2k+1)\pi/2\pm c\,\pi/(2n)$. This corresponds to an (approximate) single axis rotation with a well-controlled and reduced coupling amplitude as discussed in the preceding section. 

However, a reduced periodicity comes at a price, namely additional peaks appearing in the sensing spectra. This is best understood by noting that the filter for an $M$ times periodic repetition of an $m$-pulse block can be described as the overlap of a grating $\mathcal{G}$ with a block filter $\mathcal{F}_{\rm block}$\,\cite{ajoy13, zhao14} (see Appendix\,\ref{append_grating})
\begin{equation}\label{filtgrating} \mathcal{F}_{\pm c}(\omega)^2=\frac{1}{t^2}\,\mathcal{G}(\omega\,t,M)\,\mathcal{F}_{\rm block}(\omega\tau_0,m,c)^2\,.  \end{equation}
We will assume the number of slices $s$ to be even, in which case $M=s/2$ and $m=2n$ with a `block' consisting of a single `$c\to-c$' sequence. The grating is then given by ($m$ even) $\mathcal{G}=\sin^2(\omega t/2)/\sin^2(\omega t/(2M))$ and the block filter is calculated analogue to the CPMG filter (see Appendix\,\ref{append_deriv}) by just adapting $s(t)$ to the modified pulse sequence. Thus the total filter can be interpreted as the overlap of a grating with peak width inversely to the total time $\Delta \omega\propto1/t$ modulated by a broader block filter of width $\Delta\omega\propto (t/M)^{-1}$.  Compared to a perfectly `symmetric' grating ($M$=$N$, $m$=1), in which the grating consists of peaks $\omega\tau_0=(2k+1)\pi/2$, ($m$\,-\,$1$) additional peaks per resonance order $k$ and separated by $\Delta(\omega\tau_0)=\pi/m$ emerge for $M=N/m$, gradually introduced into the filter by reducing the symmetry of the $m$-pulse blocks\,\cite{zhao14}. 

Figure\,\ref{b_sensing}\,(b) illustrates the filters for a sequence of $\pm c$ alternating slices, a fixed total pulse number $N$=24 and different pulse numbers $n$ per slice. Previously located in the oscillatory regime  for a fully periodic $N$=24 sequence (figure\,\ref{b_sensing}\,(a)), the alternation leads to well-defined peaks of frequency width given by the inverse total time and well-described in the filter framework; though at the expense of the emergence of additional peaks for a given frequency.

\section{Conclusion}
We have shown how the evolution operator of interacting spins subject to a control pulse sequence can be described in a filter description. This holds true for (hyperfine) coupled spins without population exchange in both the limiting cases of strong and weak coupling. A universal behavior can be ascribed to the resonance region, allowing for a straightforward identification of both the interaction type and amplitude. Importantly, it has been shown that the interpretation in the filter framework can be extended to the previously inaccessible limit of long times and large rotation angles. Thus its insight can be used for the construction of tunable decoupled quantum gates. It should be noted that obtaining the exact solution of the conditional spin evolution is a straightforward task\,\cite{vandersar12, taminiau12, zhao12, kolkowitz12},  though its formulation is more involved, and quickly becomes a numerical task involving the inversion of trigonometric functions. In contrast, the filter formulation provides an intuitive approach and combining both methods can lead to further optimizations whenever the limiting cases of strong and weak coupling are not strictly fulfilled.  Moreover, we have analyzed the impact on the (control spin) subsystem coherence. Remarkably, the sensing relevant regime is well-described in the filter formulation and is closely related to the filter for classical noise. Furthermore it has been shown that sensing resolution can be improved by alternating sequences with well defined coupling amplitudes as follow directly out of the filter description.     

\begin{acknowledgements}
This work was supported by an Alexander von Humboldt Professorship, the ERC Synergy grant BioQ and the EU projects DIADEMS, SIQS and EQUAM.
\end{acknowledgements}

\appendix

\section{Limitations to the filter description}\label{append_limit}
Hamiltonian (\ref{ham1}) in a rotating frame with respect to $H_0=(\omega/2)\, \sigma_z$, well suited for the filter analysis in the weak coupling regime $A\ll\omega$, follows as $H_{\rm int}=(A/2)\,[\sigma_+ \exp(i\omega t)+\text{h.c.}]\otimes S_z(t)$.  The evolution operator in the Magnus expansion can then be written as 
\begin{equation}\label{mag1} U_{\rm int}=\exp\left(\sum_{i=1}^\infty \Omega_i(t) \right) \end{equation}
with the first two expansion contributions
\begin{equation}\label{l2} \begin{split}   \Omega_1(t)&=-i\,A/2\,\int_0^t \tilde{H}_{\rm int}(t')\,\mathrm{d}t'   \\
\Omega_2(t)&= -\frac{1}{2}\,\left(\frac{A}{2}\right)^2\int_0^t\,\mathrm{d}t_1\int_0^{t_1}\mathrm{d}t_2 \left[ \tilde{H}_{\rm int}(t_1),\tilde{H}_{\rm int}(t_2) \right]\,
\end{split}  \end{equation}
and the definition $H_{\rm int}=(A/2)\,\tilde{H}_{\rm int}$. A completely analogue treatment can be carried out for the strong coupling limit in a rotating frame with respect to the hyperfine contribution. \\A sufficient condition ensuring absolute convergence of the sum appearing in (\ref{mag1}) is given by \,\cite{casas07} $\int_0^t \lVert H_{\rm int}(t') \rVert\mathrm{d}t'<\pi$, with the matrix norm here and in the following defined as the operator norm.

As the filter description is based on the first order contribution alone, its validity is restricted to timescales on which higher order contributions are sufficiently small.
With $\lVert \tilde{H}_{\rm int}(t)  \rVert = J$, a time independent quantity as a result of the unitarily invariant matrix norm, it follows that $\lVert  \Omega_1(t) \rVert\leq (A/2) J\,t$ and $\lVert  \Omega_2(t) \rVert\leq  1/2\,(A/2)^2\,t^2\,J^2$ and under the assumption of absolute convergence\,\cite{khodjasteh08}
\begin{equation} \left\lVert\sum_{i=k}^\infty \Omega_i(t)\right\rVert\leq C_k \left(\frac{A}{2}  \right)^k(J\,t)^k  \end{equation}
with a constant $C_k=\mathcal{O}(1)$.\\
 Recalling the filter definition (see Appendix\,\ref{append_deriv}) $\mathcal{F}=(1/t)|\int_0^t\exp(i\omega t')s(t')\mathrm{d}t'|$, the first order contribution takes the form  $||\Omega_1(t)||=|A|/2 J\,\mathcal{F}\,t$  and the above estimations allow for an upper bound on time for the filter description validity. 

 More approximate, for sufficiently many pulses $N$ and close to resonance
\begin{equation}\begin{split}  \left\lVert\Omega_2(t) \right\rVert &\leq \left(\frac{A}{2}\right)^2  \int_0^t \mathrm{d}t_1 \left\lVert \tilde{H}_{\rm int}(t_1)   \right\rVert  \left\lVert  \int_0^{t_1} \mathrm{d}t_2 \tilde{H}_{\rm int}(t_2) \right\rVert \\
&\lesssim \left( \frac{A}{2} \right)^2\,J^2\,\mathcal{F}\,\int_0^t\,\mathrm{d}t_1\,t_1=\left(\frac{A}{2}J t \right)^2\frac{1}{2}\mathcal{F}
  \end{split}  \end{equation}
where the `coarse grain' estimation $\lVert \int_0^{t_1}\mathrm{d}t_2\,\tilde{H}_{\rm int}(t_2)  \rVert\lesssim \mathcal{F}J t_1$, approximately valid in the quasi-resonant regime for sufficiently long times (large $N$),  has been used. Thus, $\lVert\Omega_2(t)\rVert\lesssim ((A/2)\,F\,t\,J)^2$, which compared to the first order contribution and its associated rotation angle $\theta$   $\lVert \Omega_1(t) \rvert =(A/2 \mathcal{F}tJ)=\theta/2\,J$, reveals that the second order contribution is of $\mathcal{O}([\theta/2]^2)$. 
Thus, the general validity of the filter description is limited to small rotation angles $\theta$; except for the special cases  discussed in the main text (e.g. $c$=0 or alternating sequences). 

\begin{figure}[htb]
\begin{centering}
\includegraphics[scale=0.6]{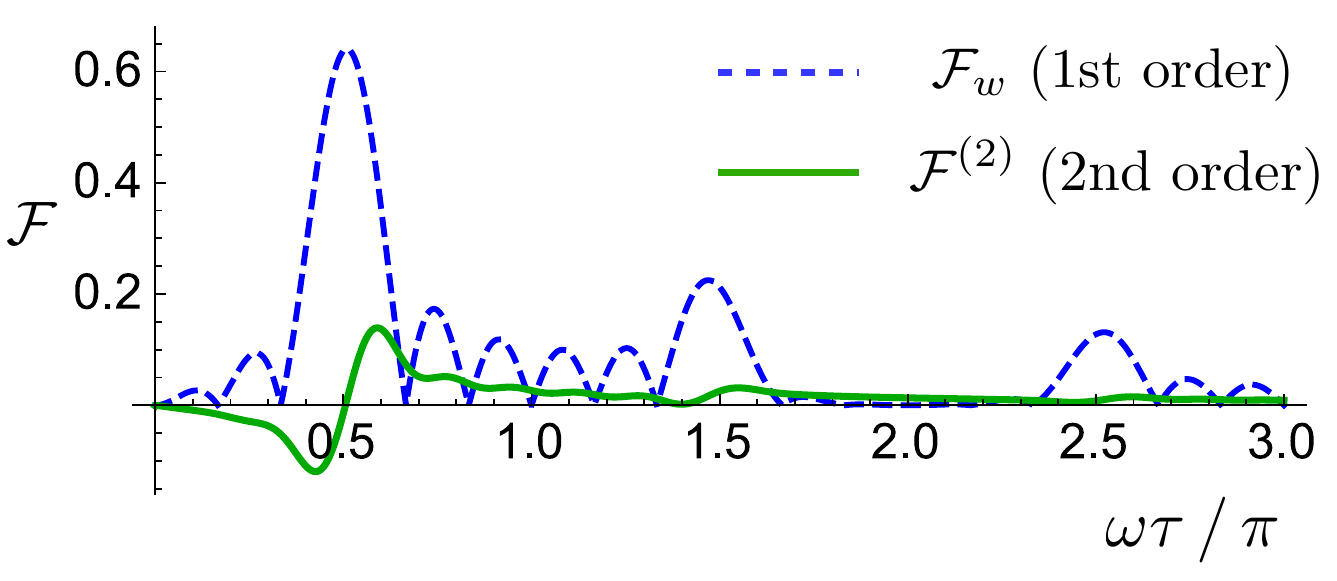}
\caption{\label{b_2ndorder} Comparison of the first and second order filter contribution in the weak-coupling limit for a number $N$=6 of periodic pulses.    }
\end{centering}
\end{figure}

In the weak coupling limit, the evolution operator (\ref{uint_wc}) up to second order takes the form $U_{\rm int}\simeq \exp(-i[\theta/2\,\sigma_\phi\otimes s_z+ (\theta/2)^2\,(\mathcal{F}^{(2)}/\mathcal{F}_w^2)\,\sigma_z\otimes \mathds{1}])$ with $\theta=A\mathcal{F}_w t$ the first order rotation angle, and $\mathcal{F}_w$ and $\mathcal{F}^{(2)}$ the first and second order filter, respectively. These filter contributions are compared in figure\,\ref{b_2ndorder}, verifying that in the resonance region $\mathcal{F}^{(2)}/\mathcal{F}_w\lesssim1$ and thus indeed the second order contribution $\lesssim (\theta/2)^2$ as expected out of the previous discussion.

\section{Filter function derivation}\label{append_deriv}
\subsection{Weak coupling limit}
The rotating frame evolution, retaining only the first order Magnus expansion contribution, takes the form 
\begin{equation} U_{\rm int}\simeq \exp\biggl(-i\frac{A}{2}\,\bigl[\sigma_+ \int_0^t\,e^{i\omega\,t'}\,s(t')\mathrm{d}t'+\text{h.c.} \bigr]\otimes s_z\biggr)\end{equation} with $s(t)$$\in$$\{-1,1 \}$ the control pulse jump function as defined in the main text.  Thus it essentially remains to calculate $\chi(t)=\int_0^t \exp(i\omega t')s(t')\mathrm{d}t'$ and the filter\,(\ref{filt_weak}) and rotation angle\,(\ref{rot_wc}) subsequently follow as $\mathcal{F}_w=(1/t)\,|\chi(t)|$ and $\tan\phi=\text{Im}(\chi(t))/\text{Re}(\chi(t))$, respectively.

Using that the pulse times for an equidistant $N$-pulse sequence are given by $t_k=-\tau+k\,(2\tau)$, with start and end time $t_0=0$ and $t_{N+1}=t\equiv 2N\tau$
\begin{equation}\label{wc_chi1}\begin{split} &\chi(t)=\sum_{k=0}^N(-1)^k\int_{t_k}^{t_{k+1}} e^{i\omega t'}\mathrm{d}t'\\
&=\frac{1}{i\omega}\left((-1)+(-1)^N e^{i\omega (2N\tau)}-2e^{-i\omega\tau}\sum_{k=1}^N\left[-e^{i\omega\,2\tau}\right]^k  \right) \end{split} \end{equation}
and upon evaluation of the sum using the geometric series property
\begin{equation}\label{chi_wc}  \chi(t)=\frac{i}{\omega}\,\left( 1-(-1)^N e^{i\omega\,t}-2\,\frac{e^{i\omega N\tau}}{\cos(\omega\tau)}\,\bfrac{-i\sin(\omega N\tau)} {\cos(\omega N\tau)}     \right)\end{equation}
with the upper and lower case corresponding to $N$ being even and odd, respectively.

\subsection{Strong coupling limit}\label{deriv_sc}
Hamiltonian (\ref{ham1}) in the rotating frame with respect to the hyperfine contribution and for the `spin-1' control spin configuration $S_z(t)=1/2(s_z(t)\pm \mathds{1})$ (corresponding to upper and lower signs in what follows)  takes the form
\begin{equation}\label{hint_sc}  H_{\rm int}=\frac{\omega}{2}\,\bigl( \varsigma_+ e^{i\,(A/2)\,[s_z\,s_f(t)\pm t ]}+\text{h.c.}  \bigr)  \end{equation}
with $\varsigma_+$ the ladder operator in the $\sigma_x$ eigenbasis, i.e. $\varsigma_++\text{h.c.}=\sigma_z$ and $i\,\varsigma_++\text{h.c.}=\sigma_y$, and $s_F(t)=\int_0^t s(t')\mathrm{d}t'$. For a periodic (CPMG) $N$-pulse sequence $s_F(t)$ can be expressed as
\begin{equation}\label{as_1} s_F(t)=\sum_{k=0}^N (-1)^k\,\left(t-k\,2\tau \right)\,\theta(t_{k+1}-t)\,\theta(t-t_k)  \end{equation}
with the pulse times $t_k=-\tau+k\,(2\tau)$ and the initial and final time $t_0$=$0$ and $t_{N+1}$=$t$, respectively.\\ A separation of (\ref{hint_sc}) into a conditional and unconditional part is obtained by noting that
\begin{equation}\begin{split} \cos\bigl([A/2] s_F(t) s_z\bigr)&=\cos\bigl([A/2]  s_F(t)\bigr) \mathds{1} \\
 \sin\bigl([A/2] s_F(t) s_z\bigr)&=\sin\bigl([A/2] s_F(t)\bigr)\, s_z \,.  \end{split} \end{equation}
The total evolution in the rotating frame and approximated by the first order Magnus expansion then follows as
\begin{equation}\footnotesize U_{\rm int}\simeq\exp\left(-i\frac{\omega}{2}\,\biggl[(\varsigma_+\,\chi^{\pm}_c+\text{h.c.}) \otimes s_z+(\varsigma_+\chi^{\pm}_u+\text{h.c.})\otimes\mathds{1}\biggr] \right)  \end{equation}
with
\begin{equation}\label{chi_sc}
\begin{split}
\chi^{\pm}_c&=i\,\int_0^t\,\mathrm{d}t' \sin\biggl(\frac{A}{2}\,s_F(t')\biggr)\,e^{\pm i\,(A/2)\,t'} \\
\chi^{\pm}_u&=\,\,\,\int_0^t\,\mathrm{d}t' \cos\biggl(\frac{A}{2}\,s_F(t')\biggr)\,e^{\pm i\,(A/2)\,t'}\,.
\end{split}
\end{equation}
The corresponding conditional and unconditional filter as defined in\,(\ref{uint_sc}) and (\ref{filter_sc}) then follow as $\mathcal{F}_{c/u}=(1/t)|\chi_{c/u}^+|=(1/t)|\chi_{c/u}^-|$, and the rotation angle as defined in (\ref{rota_sc}) is given by $\tan\phi=\text{Im}(\chi_{c/u}^+)/\text{Re}(\chi_{c/u}^+)$.

Thus it remains to evaluate the expressions (\ref{chi_sc}). Inserting the expression (\ref{as_1}) for $s_F(t)$ 
\begin{equation}\small \chi_c^{\pm}=\frac{1}{2}\sum_{k=0}^N(-1)^k \int_{t_k}^{t_{k+1}}\mathrm{d}t'\bigl(e^{i(A/2)(t'-k (2\tau))}-\text{h.c.} \bigr) e^{\pm i(A/2) t'} \end{equation}
and analogue for $\chi_u^{\pm}$. Performing the integration, inserting the expressions for $t_k$=$-\tau+k(2\tau)$, separating the `special times' $t_0$ \& $t_{N+1}=t$,  and finally using the geometric series property in close similarity to the weak coupling treatment (\ref{wc_chi1}), results in the expressions
\begin{equation}
\begin{split}
       \chi_c^{\pm}&=\left[1-(-1)^N e^{\pm i A N \tau}  \right]\,\left(\frac{i}{2A}\mp\frac{\tau}{2} \right)+e^{\pm i (A/2) N \tau}\\ &\left[\frac{i}{A}\mp\frac{\tau}{\cos\bigl((A/2)\tau\bigr)}e^{\pm i(A/2)\tau} \right] \bfrac{\pm i\sin\bigl((A/2) N\tau\bigr)}{-\cos\bigl((A/2)N\tau\bigr)}
\end{split}
\end{equation}
and 
\begin{equation}
\begin{split}
\chi_u^{\pm}&=\left[ 1- e^{\pm i A N \tau} \right] \left( \frac{\pm i}{2A}+\frac{\tau}{2}\right)+e^{\pm i(A/2)N\tau}\\ & \left[\frac{1}{A}+\frac{\tau}{\sin\bigl((A/2)\tau\bigr)}e^{\pm i(A/2)\tau}  \right] \sin\left(\frac{A}{2}N\tau\right)\,.
\end{split}
\end{equation}
Upper and lower cases in wavy brackets correspond to the total pulse number $N$ being even and odd, respectively.

\section{Spin-1/2 control spin in the strong coupling limit}\label{append_strong1h2}
For the strong coupling limit $A\gg \omega$ and assuming the control qubit to be of the `spin-1/2 type', the rotating frame Hamiltonian is given by
\begin{equation} H_{\rm int}=\frac{\omega}{2}\,\left(\varsigma_+\,e^{i\,A s_F(t)\,\,s_z}+\text{h.c.} \right)  \end{equation}
with $\varsigma_+$ and $s_F(t)=\int_0^t s(t')\mathrm{d}t'$ as defined in Appendix\,\ref{deriv_sc}.

Thus, the rotating frame evolution follows as
\begin{equation} U_{\rm int}=\exp\left(-i\frac{\omega}{2}\,t\,\left[ \mathcal{F}_c\,\sigma_y\otimes s_z +\mathcal{F}_u\,\sigma_z\otimes\mathds{1} \right] + \mathcal{O}\bigl((\omega t)^2\bigr) \right)  \end{equation}
with $\mathcal{F}_c=(1/t)\text{Im}(\zeta)$, $\mathcal{F}_u=(1/t)\text{Re}(\zeta)$ and the definition $\zeta=\int_0^t\mathrm{d}t' \exp(i A s_F(t'))$. With (\ref{as_1}) and using the same procedure as in Appendix\,\ref{append_deriv} the filters are readily evaluated to
\begin{equation}\begin{split} \mathcal{F}_c&=\bfrac{0}{1/(N A \tau) \bigl(1-\cos(A\tau)\bigr)}  \\ \mathcal{F}_u&=1/(A\tau)\,\sin(A\tau)\,.  \end{split} \end{equation}
Upper and lower cases in curly brackets correspond to an even and odd number of total pulses $N$, respectively.
Remarkably, the conditional filter $\mathcal{F}_c$ does not lead to a scalable evolution, i.e. it scales inversely with the pulse number or is even zero for an even number of pulses. More precise, at most a single pulse unit contributes to the evolution and thus such a configuration is not suitable for conditional gate interactions. In contrast, the unconditional contribution is scalable and does reveal a discrete peak structure. 

\section{Grating block-filter construction in the weak coupling limit} \label{append_grating}
The filter in accordance with the definition (\ref{uint_wc}) as follows from a first order Magnus expansion in the weak coupling limit is given by
\begin{equation}\label{grat_proof1} \mathcal{F}(t)=\left|\frac{1}{t}\,\int_0^t\,e^{i\omega\,t'}\,s(t')\,\mathrm{d}t'\right|  \end{equation}
with the step function $s(t)\in\{\pm 1\}$ as previously defined describing a pulse sequence consisting of $M$ periodic repetitions of $m$-pulse blocks. Taking into account that the time integration involved can be split into $M$ periodic time intervals $[t_k, t_{k+1}]$ with $t_1=0$, $t_{M+1}=t$ and $\Delta t=t_{k+1}-t_{k}=t/M$, allows to rewrite (\ref{grat_proof1}) as
\begin{equation}\label{grat_proof2}\begin{split} &\mathcal{F}(t)=\left|\frac{1}{t}\sum_{k=1}^{M}\int_{t_k}^{t_{k+1}} e^{i\omega t'}\,s(t')\,\mathrm{d}t'\right|\\&=\frac{1}{t}\left|\sum_{k=1}^M\,\left[ (-1)^m \right]^{k+1} e^{i\omega(k-1)\,(t/M)} \int_0^{t/M}\,e^{i\omega t'}\,s(t')\,\mathrm{d}t'\right| \end{split}\end{equation}
where in the last step the periodicity property $s(t_k+t')=[(-1)^m]^{k+1} s(t')$ has been used. Out of (\ref{grat_proof2}) and (\ref{filtgrating}) it is straightforward to identify the grating as
\begin{equation}  \mathcal{G}(\omega t,M,m)=\left| \sum_{k=1}^M\left[(-1)^m \right]^{k+1}\,e^{i\omega(k-1)\,(t/M)} \right|^2 \end{equation}
that upon evaluation of the sum using the geometric series property results in
\begin{equation}\label{grating_def} \mathcal{G}=\begin{cases} \frac{\sin^2(\omega\, t/2)}{\sin^2(\omega t/(2M))}\qquad\qquad\qquad\,\,\,\, m\text{ even} \\ \frac{1}{\cos^2(\omega t/(2M))}\,\begin{cases} \sin^2(\omega t/2)\quad m\text{ odd}, M\text{ even}\\ \cos^2(\omega t/2) \quad m\text{ odd}, M \text{ odd.} \end{cases} \end{cases}  \end{equation}
Correspondingly the `block filter' is identified as $\mathcal{F}_{\rm block}=\bigl|\int_{0}^{t/M}\exp(i\omega t')\,s(t')\mathrm{d}t'\bigr|$. For a $\pm c$ block with $n$ pulses per slice, i.e. $m=2n$, the block filter follows as $\mathcal{F}_{\rm block}^{\pm c}(2n)=|\chi_n(\tau_{+c},t=2n\tau_{+c})+(-1)^n\,\exp(i2n\omega\tau_{+c})\,\chi_n(\tau_{-c},t=2n\tau_{-c})|$ with $\chi$ as defined in\,(\ref{chi_wc}).

\bibliography{filt_design}

\end{document}